\documentclass[conference]{IEEEtran}
\usepackage{graphicx}
\usepackage{epstopdf}

\usepackage{cite}

\usepackage[cmex10]{amsmath}

\usepackage[font=footnotesize]{subfig}
\usepackage{url}

\begin{document}
\title{Capacity Based Evacuation with Dynamic Exit Signs}

\author{
\IEEEauthorblockN{Antoine Desmet and Erol Gelenbe}
\IEEEauthorblockA{Imperial College London\\
Department of Electrical and Electronic Engineering\\
Intelligent Systems and Networks Group\\
Email: \{a.desmet10, e.gelenbe\}@imperial.ac.uk}}

\maketitle

\begin{abstract}
%\boldmath
Exit paths in buildings are designed to minimise evacuation time when the building is at full capacity. We present an evacuation support system which does this regardless of the number of evacuees. The core concept is to even-out congestion in the building by diverting evacuees to less-congested paths in order to make maximal usage of all accessible routes throughout the entire evacuation process. The system issues a set of flow-optimal routes using a capacity-constrained routing algorithm which anticipates evolutions in path metrics using the concept of ``future capacity reservation". In order to direct evacuees in an intuitive manner whilst implementing the routing algorithm's scheme, we use dynamic exit signs, i.e. whose pointing direction can be controlled. To make this system practical and minimise reliance on sensors during the evacuation, we use an evacuee mobility model and make several assumptions on the characteristics of the evacuee flow. We validate this concept using simulations, and show how the underpinning assumptions may limit the system's performance, especially in low-headcount evacuations.
\end{abstract}

\IEEEpeerreviewmaketitle

\section{Introduction}
Evacuation planning is a critical step in the design process of large buildings: architects must ensure their design complies with a variety of safety regulations set by regulatory bodies. In particular, these regulations impose an upper limit on the building evacuation time: for instance, sporting venues must be evacuated within 8 to 10 minutes, regardless of their capacity \cite{FIFA_Safety_manual,Events_safety_at_Sports_Grounds}. In order to meet such constraints, architects elaborate an evacuation scheme which meets regulatory requirements when the building is at full capacity. This plan is then deemed sufficient for any other situation: the assumption is that a building filled to a lesser capacity will necessarily be evacuated in a shorter time and therefore fall within the regulatory limits.
Because of this ``worst-case" approach, buildings are only evacuated in optimal conditions when filled to full capacity, while little consideration is put into optimising evacuations when the venue is filled below capacity. Clearly, stadiums are not always filled to capacity, nor are movie theatres at every screening, or lecture theatres at every lecture, and so on. In those circumstances, it is reasonable to expect spectators will concentrate near what they regard as being the ``best seats"; or the building managers will close off parts of the viewing area when attendance is low. If an emergency evacuation is initiated in these circumstances, the exits located near the best seats will experience much higher levels of congestion and take longer to clear than exits located near other seats. This violates the ``Uniformity Principle" \cite{francis1981uniformity} which states that all exits must be used at full capacity and throughout the \emph{entire} process, for the evacuation time to be minimal.
In evacuations triggered by a fire or a bomb threat where the actual object is very difficult to find \cite{Gelenbe1997,Search,Mines,PhysRev2010}, clearly, every second counts: regardless of attendance, the area should be evacuated in minimal time. This motivates our research objective to go beyond merely satisfying the ``worst-case scenario" requirement, and instead propose a complete system which minimises building evacuation time \emph{regardless} of attendance.

\section{Background}
A review of evacuee support systems can be found in \cite{CAMWA,Wu2}.
While these solutions have merit, most use only path distance as routing metric and disregard congestion. However, congestion becomes a predominant factor when attempting to bring evacuees to safety, outside of the building, in the shortest amount of time \cite{BiDesmetGelenbeISCIS2013}. In extreme cases, the risk of stampede or uncontrolled crowd movement can become greater than the threat posed by the hazard which originally triggered the evacuation. Yet routing evacuees while managing congestion is a complex task since congestion is a sensitive metric: it increases with the probability of routing traffic into this path \cite{gelenbe2003sensible}. This is highlighted in \cite{Desm1310:Reactive}, where we implement a control system which monitors congestion and instructs evacuees to move towards a less-congested path. Another strategy is to assign routes with probabilities inversely proportional to congestion \cite{gelenbe2003sensible}. All these implementations are liable to produce oscillating routes, since the control loop ignores the sensitive nature of the routing metric. In some extreme cases, the position of the evacuees may oscillate in a back-and-forth motion if they continuously receive route correction updates which contradict the previous instruction. Oscillation damping techniques presented in \cite{network1} are effective but often require an ad-hoc parametrisation to suit the building layout, number of evacuees, etc. Chen et al. \cite{Chen2008_LoadBalancedGuide, Chen2011_Flow} also use a similar approach, with a focus on oscillation prevention.
Instead of using real-time measurements alone, some researchers acknowledge the importance of measuring and modelling the capacity of exits and access paths, which is a key aspect of congestion.
The most basic approach consists of ensuring all exits are used at their full capacity throughout the evacuation: this is Francis' fundamental ``Uniformity Principle" \cite{francis1981uniformity}. In order to achieve this, the number of evacuees assigned to any exit must be proportional to their maximum output flow. This simple concept is based on the assumption that paths leading to exits are easily reachable, have no capacity constraints, are free of interactions with other paths, and that the time needed to reach an exit is negligible compared to the time to clear the bottlenecks (exits). This theory has since been refined, in \cite{Pursals2009692} the authors use an advanced evacuee flow model where the speed of evacuees is influenced by their density.
A more thorough approach considers capacity constraints on the entire path -- not only the exit -- and that paths may ``interact" when crossing each other. In most cases, the building or area is modelled as a flow graph with restricted capacities \cite{berlin1978use,chalmet1982network} and queueing models can also be used \cite{ActaI,ActaII}. The ``Max-flow min-cut" theorem from Ford and Fulkerson \cite{ford1956maximal} identifies the minimum set of edges which provide the maximal static flow between a source and sink. While very useful to identify bottlenecks in the flow graph, this algorithm only solves the \emph{maximum dynamic flow} problem, which is to maximise flow in steady-state regime, regardless of path length or travelling times. In the context of emergency evacuations, the \emph{quickest flow} problem is more relevant: it consists of finding the paths which allow a set number of evacuees to travel from a source to an exit in the shortest amount of time. The difficulty of this problem resides in finding paths which optimally combine short distance \emph{and} large flow capacity. In order to account for transit time, the \emph{static} flow graph can be duplicated over a number of time-steps, making it a \emph{time-expanded} flow graph. By connecting nodes from different time-steps with respect to edge transit times, the time-expanded graph encodes \emph{both} flow and travel times. The quickest flow problem is usually solved by applying linear programs on the time-expanded graph \cite{Kisko1985211}. The \emph{Quickest Transhipment} extends the quickest flow problem to multiple sources and sinks and is possibly the best representation of the emergency evacuation problem. Hoppe and Tardos \cite{hoppe2000quickest_transshipment_problem} provide a literature review of the subject and also a polynomial-time algorithm for this problem. Hamacher and Tufecki \cite{NAV3220340404} propose an algorithm which both minimises evacuation time and distance covered by evacuees.
While techniques based on time-expanded graphs provide optimal solutions, the linear programs used to define routes have high computational cost: according to \cite{lu2005capacity} the solver has a complexity of $O(N)^6$, where $N$ is the number of nodes in the time-expanded graph, itself formed of $T$ duplicates of the static graph of $n$ nodes: $N=(T+1)\cdot n$. The overall computational complexity: $O((T \cdot n)^6)$ depends on the graph's complexity, and the time-horizon to solve the problem.

\section{Capacity-Reservation Algorithms}
Clearly, Linear programming approaches suffer from poor scalability due to their high computational complexity. In their paper, Lu et al. \cite{lu2003evacuation} introduce heuristic pseudo-optimal algorithms for capacity-constrained evacuation planning. Their algorithms are based on the concept of \emph{future capacity reservation}: each time a route is allocated to an evacuee, an algorithm reserves capacity for this individual on each node at the expected time of arrival. Capacity reservations are made by decreasing the edge's capacity associated to the \emph{time-step} which covers the expected arrival time. If an evacuee is scheduled to arrive at a time where all the node's capacity has already been reserved, the system assumes the evacuee will be held there until such time as some capacity becomes available again. This method effectively builds a forecast of the congestion in the building, which is updated at each route assignment. Subsequent path assignments are made using a modified version of Dijkstra's shortest path algorithm, which calculates path traversal times based on the congestion forecast. The capacity of edges for each time slice is stored in a time-series format; we can further reduce the algorithm's complexity by increasing the time-step duration, at the cost of path optimality, since this also decreases the time resolution.
The authors combine these features into CCRP, a routing algorithm with a computational complexity of $O(p\cdot n\cdot log(n))$ where $p$ is the number of evacuees and $n$ the number of nodes. The complexities of this algorithm and linear programming methods have different parameters and cannot be compared analytically, however the author's result indicate at least a threefold reduction in algorithm run-time.
The main drawback of CCRP is its use of Dijkstra's shortest path algorithm, which performs an entire search of the network at each step.

\subsection{Cognitive Packet Network Routing Algorithm}
These limitations have lead us to replace CCRP's Dijkstra shortest-path algorithm with CPN (Cognitive Packet Network), a self-aware routing algorithm. Unlike Dijkstra's shortest path algorithm which performs full-graph searches, CPN uses Random Neural Networks (RNN) to reduce the overhead associated with route discovery or route maintenance. Each node in the network issues ``Smart Packets" to discover and update path metrics. Initially, these packets explore the network randomly in search for exit paths. Once an exit path is found, an acknowledgement packet travels backwards along the same route and provides feedback to the RNNs. The trained RNNs are then able to guide the subsequent Smart Packets on a hop-by-hop basis towards regions of the network which are perceived as most worthwhile -- thus cutting down on random graph exploration. The motion of the Smart Packets retains a small element of randomness, needed to maintain a set of auxiliary routes and prevent RNN overtraining. Each node maintains its own list of routes to the exits using information provided by Smart Packet acknowledgements, thus CPN performs source-routing.
Our proposed capacity-constrained routing algorithm retains CCRP's original path-delay metric based on future capacity reservations. A stream of Smart Packets is sent throughout the route allocation process to monitor changes in the routes, as capacity decreases after each route assignment through the reservation process. Because we use a model to predict future path metrics, the algorithm can operate off-line without any posterior corrections as long as the congestion prediction model is accurate. CPN will run only once, at the beginning of the evacuation, and only requires the initial distribution of evacuees to commence execution. The distribution of evacuees at the beginning of the evacuation process can be estimated using vision-based techniques in stadiums, or a summary of computers currently in use in an office environment, or connections to the local wireless access points in public buildings, etc.
Since CPN is originally a data network routing protocol designed to run on every node of a network, it is well suited for distributed deployment. As each node maintains its own routing table, the network as a whole is resistant to denial of service atacks and localised failures \cite{DoS,sakellari2010demonstrating}. It is also decentralised: Smart Packets are guided through a collective effort by each node, and in return the information they gather is shared among every node visited along the path.
Let us propose a deployment scheme in a building featuring a dense array of networked nodes. Each of these nodes can be a CPN node in addition to managing their local capacity reservations. This type of deployment provides scalability and resilience, and is well suited to the use of dynamic exit signs which we introduce later in this paper.
\subsection{Experimental Results}
\begin{figure}[htb]
\centering
%[ trim l b r t ]  (Left Bottom Right Top)
\includegraphics[width=0.45\textwidth]{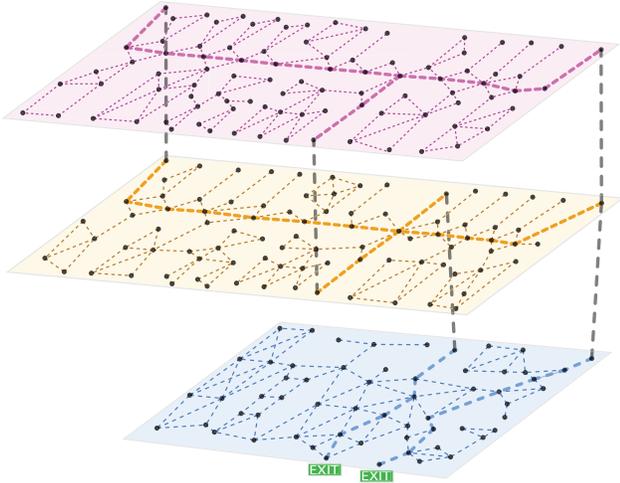}
\caption{3-D Graph representation of the building model. The building features two exits on the first floor which are marked by green signs.}
\label{fig: graph3D}
\end{figure}

\begin{figure}[htb]
\centering
%[ trim l b r t ]  (Left Bottom Right Top)
\includegraphics[width=0.45\textwidth]{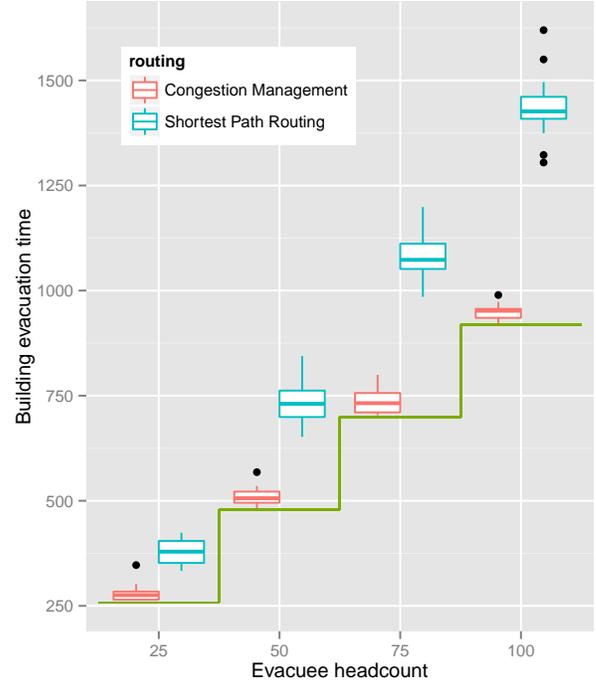}
\caption{Evacuation times using our proposed algorithm and the Shortest Path routing. We use box-plots to represent the 20 samples: the top and bottom ``whiskers" cover the top and bottom quarters of the samples, the box covers the ``central" 50\% of the samples, with the median marked by a line. The green line indicates the lower-bound of the evacuation time.}
\label{fig: routing_perf}
\end{figure}

We use a dedicated building evacuation simulator (DBES) \cite{Dimakis2010_EvacuationSimulator} to evaluate the performance of CPN as a capacity-constrained routing algorithm for evacuees. The featured graph (Figure \ref{fig: graph3D}) represents the three lower floors of Imperial College London's EEE building, where each floor has a surface area of approximately 1000~m\textsuperscript{2}. The building combines office and classroom space with a large lobby area on the ground floor, where the two exits are located.
The graph representation of this building has approximately 250 nodes and 400 edges. In order to make the evacuation non-trivial, all evacuees are concentrated on the first floor: under shortest-path routing conditions, this would lead to an over-usage of the building's central staircase, while the eastern staircase would be virtually empty. This scenario challenges the capacity-constrained routing algorithm: to achieve the quickest flow, a precise amount of evacuees must be diverted to the eastern staircase.
We use Chen and Hung's formula \cite{Chen1993125}, to get a lower bound on evacuation time of $n$ individuals through one of the staircases: $T_P(n) = T_P(1) + (n-1)t_{max}$, the transmission time of $n$ units through a path $P$ equals to the ``lead-time" of $P$ and the time to clear $n-1$ units through the path's bottleneck, i.e. the edge with the highest transit time $t_{max}$. In accordance with the ``Uniformity Principle", we distribute evacuees evenly across each staircase: indeed both staircases can be considered equally accessible from the first floor, and have identical flow characteristics. The lower-bound evacuation time appears in green on Figure \ref{fig: routing_perf}. This Figure
also shows the simulated building evacuation times for our proposed routing algorithm and for evacuations where users simply follow the shortest path to an exit. We run 20 simulations for each configuration, with randomised evacuee departure points (all on the first floor). The figure shows how an uneven distribution of evacuees can greatly decrease the flow-efficiency of shortest-path routes. In contrast, our proposed algorithm reaches a near-optimal solution in spite of the uneven evacuee distribution, and the span of the box-plots confirms that the performance is highly predictable.
\section{Dynamic Signs}
While there is a large body of research on evacuee routing algorithms, research dealing with the means to inform evacuees of these routes are, comparatively, scarce. In most cases, possession of a personal handheld communication device, like a smartphone, is assumed. This is a compelling and convenient solution especially since CPN performs source-routing: the entire route can be downloaded and displayed on the user's device screen. However, we argue that the use of such devices is unpractical for reasons: it implies device ownership, compatibility, prior installation of the application, good state of charge, remembering to consult the device in an emergency, etc. Most importantly, watching the device's screen while making one's way out of a building in the presence of large crowds and smoke or fire is hazardous in itself: evacuees should instead be ``watching their step" and focus on their environment to avoid tripping or getting crushed. Worse still, an evacuee trying to pick up his device from the ground after dropping it could easily be trampled upon and start a stampede.
\subsection{Dynamic Exit Signs}
Instead of personal communication devices, we consider the use of exit signs: they are a common feature of buildings worldwide, which users are accustomed to. As exit signs are integrated into the environment, evacuees are likely to notice them while looking for a way out. In particular, we consider \emph{dynamic} exit signs, whose pointing direction can be modified at any time. The concept of dynamic exit signs is relatively new and has a large potential, despite receiving limited research attention \cite{Roya2013DynamicSign} and being fitted in very few buildings.
\subsection{Sign Direction Scheduling Algorithm}
Exit signs, by nature, are only suited to hop-by-hop routing, therefore our main challenge resides in decomposing the complete routes issued on an individual basis to evacuees by the routing algorithm into a set of ``hops" which can be displayed by each dynamic exit sign along the way. Possibly the most straightforward method is to identify which evacuee stands in front of the exit sign, and display the corresponding direction on the sign. This is impractical for many reasons: it requires identification and tracking of evacuees, and also unrealistic: it is practically impossible to display different directions to evacuee arriving in a group.
Because our model considers only one type of evacuee, a route which is fit for an evacuee is also suitable for any other evacuee. Since we assume all evacuees have approximately the same speed, we could exchange routes between two evacuees as they walk past each other: the congestion in the building would remain unchanged, and the flow-based routing solution would not be violated. This means that reassigning a route to other evacuees along the way has no consequences, as long as the exchange is made between evacuees present at the same time at the same location. If the time-steps are sufficiently small, we can relax the ``same  time" requirement to the span of a time-step. Likewise, if the building graph is dense enough, we can relax the ``same location" requirement to the area covered by a node. Evacuee identification is no longer required, since paths can be arbitrarily reassigned within the same time-step and node.
We foresee problems if the direction displayed by signs changes too often: an evacuee who sees a sign pointing in several directions as he walks past it will either be confused or dismiss the advice, considering the sign as faulty. To improve the system's effectiveness and user-friendliness, we must reduce the rate at which the sign changes direction. We can do this by grouping next-hop directions and assigning them based on the order of arrival of evacuees. Let us explain this with a practical example: 12 capacity reservations have been made to a node in a given time-step. Out of the 12 corresponding routes, 6 continue with a left-turn, while the remaining 6 take a right turn. Instead of alternating the sign's direction between left and right each time an evacuee walks past, the sign can direct the first 6 evacuees towards the left, and the last 6 evacuees towards the right, thus minimising the number of times the sign changes directions within the time-step.
Finally, to avoid deploying a set of sensors to detect the presence of evacuees in front of the sign, we estimate the passage of evacuees using elapsed time and the mean arrival rate. We derive the time-step's mean arrival rate using the number of reservations made in that time-step, and if we assume the \emph{real} arrival rate is steady, it is indeed equal to its mean. In practice, this consists of displaying a sign for a duration proportional to the number of evacuee we wish to affect. Continuing on our previous example, instead of detecting the presence of the first and last 6 evacuees (to decide when to change direction), the system displays the first direction for $6/12=50\%$ of the time-step, and the other direction for the remaining time, assuming evacuees arrive at a regular rate.
This method breaks down individually-assigned, source-routed paths into a schedule of ``next-hop" direction to be displayed by dynamic exit signs over the course of the evacuation. This rather elegant solution has the advantage of requiring no additional sensors, however, it relies heavily on these assumptions:
\begin{itemize}
  \item The evacuee motion model is accurate: there is only one broad class of users and actual walking speeds are narrowly and evenly distributed around a well-estimated mean,
  \item The flow of evacuees arriving to any node can be considered constant and invariant over a time-step,
  \item Grouping users by directions within a time-step does not invalidate the model used to build the congestion-optimised routes.
\end{itemize}
This summary of assumptions reveals at least two parameters which are likely to affect the system's performance: the time-step's duration, and the accuracy of the motion model.

\subsection{Experimental results}
\begin{figure}[htb]
\centering
%[ trim l b r t ]  (Left Bottom Right Top)
\includegraphics[width=0.49\textwidth]{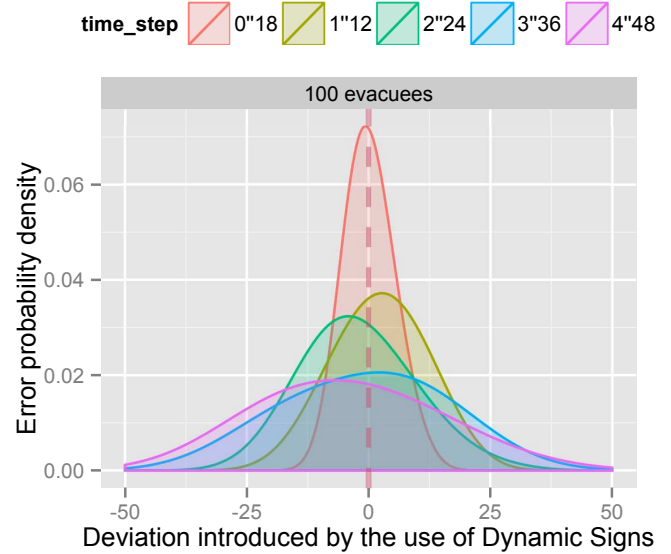}
\caption{Density of error introduced by the dynamic signs: bias added to the routing algorithm's original distribution of evacuees on staircases in percentage points.}
\label{fig: error_density}
\end{figure}

We simulate the system with varying evacuee headcount and time-step durations. While we recognise the importance of conducting a sensitivity analysis of the motion model, we leave this for the next step of our research. We assume the evacuees follow the advice which is displayed by the sign at the moment they walk past it, and do not model evacuee behavioural factors such as the decision to follow a group of evacuees regardless of the signs' advice.
\begin{figure*}[t]
\centering
\includegraphics[width=\textwidth]{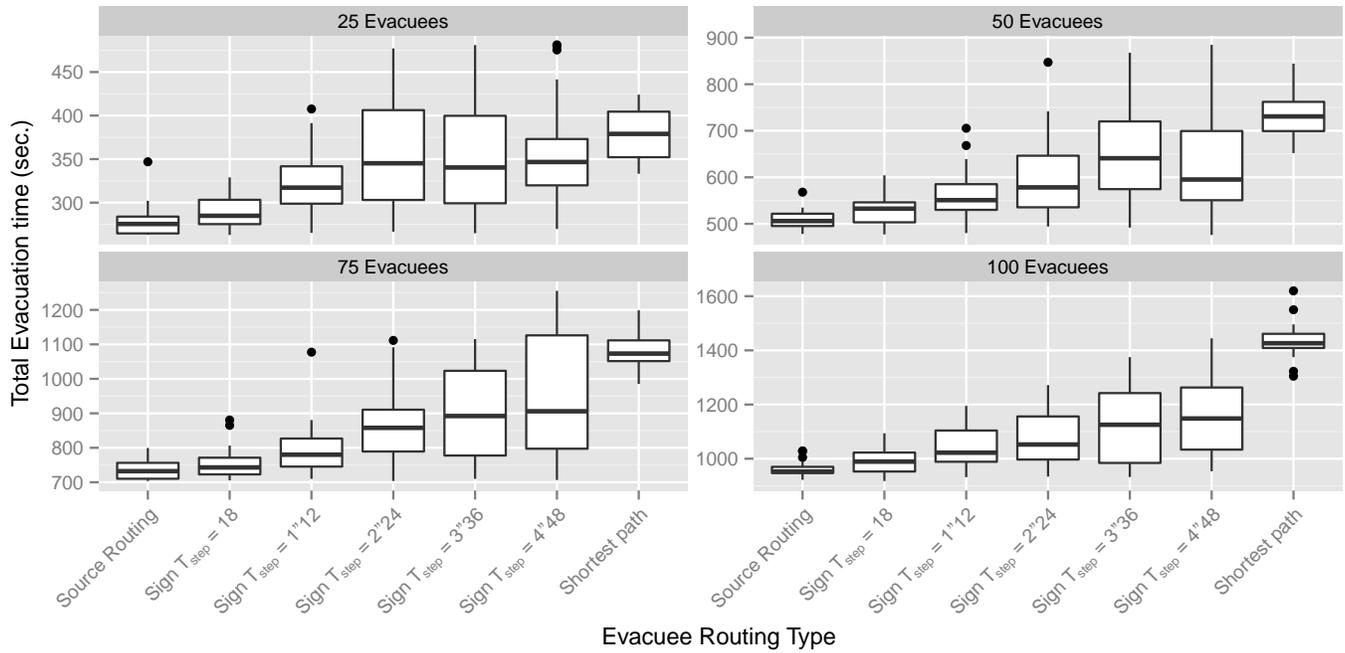}
\caption{Building evacuation times. The leftmost box-plot shows the results from the previous section, i.e. without the use of signs, while the rightmost box-plot shows results of simulation where evacuees follow the shortest path.}
\label{fig: evacTimes}
\end{figure*}
Figure \ref{fig: evacTimes} shows the building evacuation times obtained from 25 simulation runs, with randomised evacuee starting positions (all on the first floor). The left- and rightmost box-plots are carried over from the previous section's experiments: the left one shows results without dynamic signs, and the rightmost one shows shortest-path evacuation results. Let us start our analysis of Figure \ref{fig: evacTimes} with the evacuation featuring 100 evacuees. We see a clear trend where the evacuation time and spread increase with the time-step duration. This is because the steady-flow assumption -- upon which the system is based -- is progressively invalidated as the time-step duration increases. Indeed, increasing time-step duration decreases the system's resolution and allows smaller variations in arrival rate to be unaccounted for, which means directions are displayed to an increasingly \emph{approximative} number of evacuees. As a result, the paths taken by evacuees progressively diverge from the routing algorithm's flow-optimal paths, and the evacuation time mechanically increases. In contrast, smaller time-steps are better suited to ``track" the variations in arrival rate, and the system is able to precisely coordinate the display of signs with the \emph{true} arrival of evacuees.
In order to verify this, we isolate and measure the error introduced by the sign-driving algorithm: a sample of the results is on Figure \ref{fig: error_density}. Measuring the \emph{overall} error introduced by the signs is difficult, so we focus on the most critical area of the building graph: the two staircases leading to the ground floor. Let us recall that these staircases are the building's main bottlenecks, they are easily accessible to any evacuee, operate in parallel and have identical capacity, therefore under the ``Uniformity Principle" evacuees should be distributed evenly across them, i.e. a 50\%-50\% ratio. At the beginning of each simulation, we determine this assignment ratio from the routes issued by routing algorithm. At the end of the simulation, we measure the ratio which was realised by the signs. The difference between both ratios, in percentage points, is a partial measure of the error introduced by the dynamic signs. Figure \ref{fig: error_density} shows an empirical probability density of this error. The distributions associated with small time-steps are centred on 0\% and narrow, which confirm this setting maximises the signs' effectiveness in implementing whatever the routing algorithm's plan are. As the cycle time increases, the distributions become wider and flatter, which indicates that the signs gradually introduce a bias to the routing algorithm's original plans. Figure \ref{fig: error_density} shows results for 100 evacuees, but the same trend appears regardless of the evacuee headcount.
A second trend is also visible, on the evacuations featuring 25-50 evacuees: the evacuation times tend to ``plateau" beyond a certain time-step parameter, which depends on the number of evacuees. This is because the time-step has become much longer than the time it takes for evacuees to walk past the signs, thus breaking the algorithm's fundamental steady-flow assumption. In evacuations featuring 25 evacuees, it only takes them 70 seconds to clear the first floor. The results where the time-step = 2"24 (144 sec.) are poor because one time-step covers nearly twice the time it takes for evacuees to vacate the first floor. While the signs which direct evacuees to one or another stairway operate under the assumption that evacuees arrive at a steady rate throughout the entire 144 sec. of the time-step, the reality is, in fact, far from this since the flow of evacuees dries-up from 70 seconds onwards. To illustrate the effects, let us imagine a sign programmed to direct evacuees towards the central staircase during the first half of the time-step (72sec), then towards the eastern one during the second part, in order to achieve a 50\%-50\% assignment ratio towards each staircase. If evacuees only walk past the sign for the first 70 sec. of the time-step, they will all be directed towards the first staircase, and none will be towards the second one. This explains why we can observe worse performance levels than shortest-path routing. However, because signs are independent and determine randomly the order in which the sequence of directions is displayed, the system may, by chance, achieve a good balance, as much as it may completely fail to do so, which explains the wide distribution of results and indicates that the system does not perform predictably in these conditions.
\section{Conclusion}
We have introduced a system which redirects evacuees in a way that reduces the overall congestion in a building and minimises its evacuation time. Users are guided using intuitive dynamic exit signs, which are controlled by a capacity-constrained routing algorithm. The system only requires the initial distribution of users to operate, which makes it robust to component failure during the evacuation. However, this reduced dependance on sensors is somewhat offset by a heavy reliance on an evacuee mobility model, and other underpinning assumptions.
We have demonstrated that our system's performance is influenced by the duration of the system's time-step. On one hand, the dynamic signs must be able to go through a few time-steps before all evacuees have vacated critical bottlenecks in order to implement the routing algorithm's solution. On the other hand, evacuees may be confused if signs change directions too often, which is a side-effect of reducing the time-step. We are clearly in the presence of an optimisation problem, and without any research available on the response of evacuees to dynamic signs, we recommend taking a conservative approach: set the slowest possible time-step while preserving an \emph{acceptable} level of routing accuracy.
The fact that the system must go through a few time-steps to perform acceptably means that, for a given time-step duration, larger crowds will produce better results: their evacuation takes a longer amount of time, during which the system can perform more time-steps. This is a desirable feature: evacuations featuring a large number of evacuees inherently pose higher risks. However, our initial objective was to provide a system which minimises evacuation times \emph{regardless} of the number or distribution of  building occupants. Maintaining performance in smaller evacuee headcount situations requires ever shorter time-steps: this may be a limiting factor for this system.

\end{document}